\documentclass[apj,twocolumn]{emulateapj_mod}

\newcommand{\gapprox}{\,\rlap{\lower 2.5pt 
\hbox{$\sim$}}\raise 1.5pt\hbox{$>$}\,}
\newcommand{\lapprox}{\,\rlap{\lower 2.5pt 
\hbox{$\sim$}}\raise 1.5pt\hbox{$<$}\,}
\newcommand{\msun}{{M_{\odot}}}
\newcommand{\zsun}{{Z_{\odot}}}

\newcommand{\chiquad}{\ensuremath{\chi^{2}~}}
\newcommand{\chiquadr}{\ensuremath{\chi^{2}_{\rm r}}}
\newcommand{\zh}{[{{\rm Z}/{\rm H}}]}

\newcommand{\ttru}{\ensuremath{t_{\rm trunc}}}
\newcommand{\mguv}{\ensuremath{{\rm Mg}_{\rm UV}}}
\newcommand{\mstar}{\ensuremath{M^{*}~}}
\newcommand{\microjan}{{$\mu Jy$}}
\slugcomment{Submitted Version}
\shortauthors{Maraston et al.}
\begin{document}
\bibliographystyle{apj} 
\title{Evidence for TP-AGB stars in high redshift galaxies, and their effect on deriving stellar population parameters}
\author{
         C. Maraston \altaffilmark{1},
	 E. Daddi \altaffilmark{2,7},
	 A. Renzini \altaffilmark{3},
	 A. Cimatti \altaffilmark{4},
	 M. Dickinson \altaffilmark{2},
	 C. Papovich \altaffilmark{2},
	 A. Pasquali \altaffilmark{5},
	 N. Pirzkal~\altaffilmark{6}
}
\altaffiltext{1}{Oxford University, Denys Wilkinson Building, Keble Road, OX1 3RH, Oxford, UK} \email{maraston@astro.ox.ac.uk}
\altaffiltext{2}{National Optical Astronomy Observatory, 950 N. Cherry Ave., Tucson, AZ, 85719}
\altaffiltext{3}{INAF-Osservatorio Astronomico, Vicolo dell'Osservatorio 5, I-35122 Padova, Italy}
\altaffiltext{4}{INAF-Osservatorio Astrofisico di Arcetri, L.go E. Fermi 5, I-50125, Firenze, Italy}
\altaffiltext{5}{Max-Planck-Institute for Astronomy, Konigstuhl 17, 69117 Heidelberg, Germany}
\altaffiltext{6}{Space Telescope Science Institute, 3700 San Martin Drive Baltimore, MD 21218 USA}
\altaffiltext{7}{{\em Spitzer} Fellow}
\begin{abstract}
We explore the effects of stellar population models on estimating star
formation histories, ages and masses of high redshift galaxies. The
focus is on the Thermally-Pulsing Asymptotic Giant Branch (TP-AGB)
phase of stellar evolution, whose treatment is a source of major
discrepancy among different evolutionary population synthesis. In
particular, besides the models usually adopted in the literature, we
use models (by Maraston 2005), in which the contribution of the TP-AGB
phase is calibrated with local stellar populations and is the dominant
source of bolometric and near-IR energy for stellar populations in the
age range 0.2 to 2 Gyr. These models also have an underlying different
treatment of convective overshooting and Red Giant Branch stars.  For
our experiment we use a sample of high-$z$ ($1.4 \lapprox z \lapprox
2.5$) galaxies in the Hubble Ultra Deep Field held to be mostly in
passive evolution, with low-resolution UV-spectroscopy and
spectroscopic redshifts from GRAPES, and {\it Spitzer} IRAC and MIPS
photometry from the Great Observatories Origins Deep Survey.  We
choose these galaxies because their mid-UV spectra exhibit features
typical of A- or F-type stars, therefore TP-AGB stars ought to be
expected in post-Main Sequence.  We find that indeed the TP-AGB phase
plays a key role in the interpretation of {\it Spitzer} data for
high-$z$ galaxies, when the rest-frame near-IR is sampled.  When
fitting without dust reddening, the models with the
empirically-calibrated TP-AGB phase always reproduce better the
observed spectral energy distributions (SEDs), in terms of a
considerably smaller \chiquad\/.  Allowing for dust reddening improves
the fits with literature models in some cases. In both cases, the
results from Maraston models imply younger ages by factors up to 6 and
lower stellar masses (by $\sim 60 \%$ on average). The observed
strengths of the \mguv\/ spectral feature compare better to the
predicted ones in the case of the Maraston models, implying a better
overall consistency of SED fitting. Finally, we find that photometric
redshifts improve significantly using these models on the SEDs
extending over the IRAC bands.  These results are primarily the
consequence of the treatment of the TP-AGB phase in the Maraston
models, which produces models with redder rest-frame optical to
near-IR colors. This work provides the first direct evidence of TP-AGB
stars in the primeval Universe.
\end{abstract}
\section{Introduction}
Age-dating the stellar populations of galaxies provides us with a
cosmic timescale which is independent of cosmological models, and with
a mean of reconstructing their star-formation history. Moreover, the
same population synthesis tools used for age dating provide also an
estimate of the stellar mass of galaxies. Thus, deriving star
formation histories and masses for large numbers of galaxies at
various redshifts allow us to attempt an empirical reconstruction of
the formation epochs and mass-assembly history of
galaxies. Theoretical models for galaxy formation and evolution can
then be adjusted in order to comply with the galaxy evolution scenario
emerging from the observations.

Clearly, the evolutionary population synthesis tools play a critical
role in this process, and we must pay as much attention as possible to
ensure that they provide ages and masses as accurate as possible. In
this paper we focus on the problem of age and mass determinations for
galaxies which are dominated by stars in the age range $0.2 \lapprox
t/{\rm Gyr}\lapprox 2$ Gyr, because this is a range where different
population synthesis tools give quite discrepant results. Moreover, at
redshifts beyond 2 (or 3) the universe itself is younger than $\sim 3$
(or $\sim 2$) Gyr, and galaxies at such high redshift are bound to
contain stars younger than $\sim 2$ Gyr. Hence, the use of an accurate
stellar population tool is essential for the correct interpretation of
the properties of high redshift galaxies, right at a cosmic time when
different renditions of the hierarchical galaxy formation paradigm
diverge dramatically \citep[see e.g., Fig. 9 in][]{fonetal04}.

In stellar populations with ages $0.2 \lapprox t/{\rm Gyr}\lapprox 2$
Gyr, stars in the advanced evolutionary stage known as the
thermally-pulsing asymptotic giant branch (TP-AGB) contribute up to
$\sim 40\%$ of the bolometric light, and up to $\sim 80\%$ of the
near-IR \citep[][hereafter M98]{mar98}. The first appearance of TP-AGB
stars in the life of a stellar population has been called the AGB-{\it
phase transition} \citep{renbuz86}, and its direct effect is a strong
increase of the near-IR flux, as TP-AGB stars emit most of their
radiation at $\lambda >1\mu$m.  With the advent of the Spitzer Space
Telescope (SST), the rest-frame near-IR is now accessible to very
high-redshifts, which allows one an improved analysis of the spectral
energy distribution (SED) of high redshift galaxies, and makes it even
more important than before to dispose of reliable stellar population
tools.

The following concrete example illustrates the case.  For a sample of
high-$z$ galaxies with photometric redshifts $<\!z\!>\sim 2.4$ and
with data extending from the rest-frame UV to near-IR (from SST) Yan
et al. (2004) estimated an age of 1.5 to 3.5 Gyr for the dominant
stellar population ($\sim 99\%$ by mass), with the residual trace of
young stars as due to a recent star-formation episode. The Yan et
al. result was obtained using the synthetic population tool by
\citet[][hereafter BC03]{brucha03}. Some of the same galaxies were
re-analysed by \citet[][hereafter M05]{mar05} using her own synthetic
stellar populations, obtaining ages $\sim 0.6-0.8$ Gyr, right in the
middle of the epoch when TP-AGB stars dominate. Moreover, the derived
stellar masses are smaller when using Maraston models ($\lapprox
10^{10} \msun$ versus the typical value of $\sim 10^{11} \msun$ quoted
by Yan et al. 2004). Whereas both results may be astrophysically
plausible, worth emphasising is that their implications concerning the
formation redshift for the bulk of the stars and the galaxy mass
assembly are vastly different.  The difference is likely the result of
the different weight given to the TP-AGB in the two sets of models,
with the TP-AGB contribution being systematically lower in the BC03
models compared to the models of Maraston (2005). When using BC03
models a massive contribution of old stars is required in order to
match the observed near-IR fluxes, and in the meantime a sprinkle of
young stars is necessary to account for the rest-frame near-UV. On the
contrary, a two-component stellar population is not required by the
M05 models, as a single age (younger) population accounts at once for
the SED from the near-UV to the near-IR. It is interesting to note in
this context the early result of \citet{lil87}, who found for a sample
of $z\sim 0.45$ cluster galaxies that synthetic near-IR colours not
including the AGB phase were not red enough to match the data at given
optical colour. The problem was cured by adding AGB effects on top of
these early models using the fuel consumption theorem.

Besides the treatment of the TP-AGB, the BC03 and M05 models also
differ for the set of stellar isochrones used to construct the
synthetic populations, which may account for part of the difference in
the results.

In this paper we investigate further on the effect of
stellar population recipes on the modeling of galaxy SEDs, and we do
so by applying the two sets of models to infer the mass and age (or
best-fit star-formation histories) to a specific sample of
high-redshift galaxies. Note that the results obtained with the BC03
models are representative of what would be obtained with other widely
used models, like P\'egase \citep[]{fioroc97} and Starburst99
\citep[]{vazlei05} due to similar recipes for the TP-AGB phase and
identical input stellar evolutionary tracks (M05).

In Section 2 we recall the main differences between the BC03 and the
M05 models, whereas Section 3 describes the main observational
characteristics of the galaxies selected for this study.  In
Section 4 we describe the fitting procedure used to estimate the
properties of these galaxies and the results are presented and
discussed in Section 5. Finally, Section 6 is dedicated to a general
discussion of the main results of this investigation. We adopt the
current {\it concordance cosmology}, with $\Omega_{\Lambda}$,
$\Omega_{M}$ and $h~(=H_{0}[{\rm km s^{-1} Mpc^{-1}}/100])$ equal to
0.7, 0.3 and 0.7, respectively. The age of the best-fit model is
required to be lower than the age of the universe at the given
spectroscopic redshift.
\section{Main differences between the tested stellar population models}
The BC03 and the M05 stellar population models differ by the following
aspects: i) the stellar evolutionary models used to construct the
isochrones; ii) the treatment of the TP-AGB phase; and iii) the
procedure used for computing the integrated spectra. This latter point
will not be discussed separately here (see M05). For full details on
what below, we refer the reader to M05.

{\bf Input stellar models}. The BC03 models are based on the Padova
stellar tracks \citep[e.g.,][]{fagetal94}, whereas the M05 models are
mostly based on the Frascati stellar tracks
\citep[e.g.,][]{casetal97}. The two set differ mostly for: a) the
Padova tracks include a certain amount of convective overshooting on
the Main Sequence (MS), whereas the Frascati tracks were constructed
assuming no overshooting; b) the temperature distribution of the red
giant branch (RGB) phase, that is shifted to cooler temperatures in
the Padova tracks for solar metallicity and above, a result of the
different calibration of the mixing-length theory used for the
modeling of envelope convection.

Convective-core overshooting has two effects on stellar models. First,
it prolongs the MS lifetime for given stellar mass, while leaving the
effective temperature almost unchanged. This results from the extended
convective core ensuring more hydrogen fuel to the central regions
where nuclear burning takes place. Thus, for given age of the MS
turnoff, isochrones constructed from models with overshooting are
hotter, therefore bluer, than those constructed from models without
overshooting. Second, overshooting delays the appearance of stars with
a degenerate helium core, and the accompanying development of the RGB.
For example, the RGB develops at an age of $t\sim 0.5$ Gyr in the
Frascati models, and at an age of $t\gapprox 1$ Gyr in the Padova
models \citep[][M05]{feretal04}. The development of the RGB makes the
SED redder, so in the age range $\sim 0.5-1$ Gyr models with
overshooting are bluer, as they lack a well-developed RGB.

The net effect is that theoretical SEDs based on overshooting models
are bluer therefore give systematically older ages compared to those
based on models without it. Furthermore, by having a reduced
contribution by the RGB, population models using stellar tracks with
overshooting have lower luminosity for given age and mass in stars,
i.e., have higher $M/L$ ratios, hence higher galaxy masses are derived
for given luminosity. In summary, the net effect of overshooting is
that for a given SED the BC03 models indicate older ages and higher
masses compared to M05 models. Of course, this applies to population
ages such that MS stars do have a convective core, i.e., for ages
younger than a few Gyr.

While it is likely that overshooting makes convective cores somewhat
more extended in mass, the size of this extension remains somewhat
conjectural. On the one hand there exists no firm prediction from
first principles, like in most situations involving turbulence. For
stellar populations between $\sim 1$ and $\sim 10$ Gyr, the
corresponding stellar masses at the MS turnoff range from $\sim 2$ to
$\sim 1\,\msun$, with the mass of the convective core decreasing with
decreasing mass, and eventually vanishing for $M\sim 1.2\,\msun$, the
precise value depending on composition. Thus, what one would need to
know is the amount of convective overshooting (e.g., expressed as a
fraction of the mass of the convective core) as a function of the
stellar mass.  This unknown function can only be empirically
constrained by observations.  In a recent attempt, based on fitting
the shape of the color-magnitude of some open clusters,
\citet{vanetal06} concluded that overshooting would amount to $\sim
10\%$ of a pressure scale height, of course vanishing as the
convective core itself vanishes. An overshooting about this size is
also indicated by the asteroseismology of a $\beta$ Cep star
\citep{dupetal04}, and by and large is consistent with the amount
assumed in the Padova models. On the other hand, observations of LMC
star clusters rule out the late RGB development implied by the Padova
tracks \citep{feretal04}. To decide what is the size of overshooting
as a function of stellar mass goes well beyond the aims of this paper,
and we shall limit ourself to exploring the differences in the results
when using different sets of models.

{\bf The Temperature of the RGB}. Whereas for ages less than $\sim 1$
Gyr the M05 code gives redder SEDs compared to the BC03 code, after
the completion of the RGB phase transition this tendency is reversed
and the MO5 SEDs are bluer (Figure 27 in M05). This is due to the
warmer RGBs in the Frascati stellar tracks compared to the Padova ones
(Figure~9 in M05), which in turn results from different calibrations
of the mixing-length parameter used in modeling envelope
convection. The ages derived with the M05 models are {\it older} than
those obtained from the BC03 models.

{\bf The TP-AGB phase}. The treatment of the TP-AGB phase is perhaps
the main difference between the M05 and the BC03 models. The
contribution of the TP-AGB stars to the integrated light of a
synthetic stellar population critically depends on what is adopted for
the stellar mass loss taking place during this phase.  The higher the
assumed mass loss, the sooner the star loses its envelope, and the
sooner the TP-AGB phase is terminated. Once more, there is no theory
relating the mass loss rate to the basic stellar parameters, and once
more one has to rely on empirical calibrations. BC03 adopt the mass loss
calibration by \citet{vasetal93}, tuned to reproduce the maximum
TP-AGB luminosity of a sample of Magellanic Cloud clusters from
\citet{froetal90}. In M05 instead, what is calibrated is the
fractional contribution of the TP-AGB to the total bolometric light,
and this calibration is made essentially on the same MC clusters using
the Frogel et al. data. Sharing basically the same observational
calibrating data, the two codes should give consistent results. As
shown in M05 they do not, and the effective contribution of the TP-AGB
is much larger in M05 compared to BC03. Full illustration of the
differences is given in M05, where it is also shown that other
population synthesis models, namely P\'egase and Starburst99 behave
very much like the BC03 ones.  Again, rather than tracing the origin
of the discrepancy, we focus here on the effects on the study of high
redshift galaxies.

Summing up the effects of both overshooting and TP-AGB, the M05 models
are {\it brighter} and {\it redder} than the BC03 models for ages
between $\sim 0.2$ and $\sim 2$ Gyr. Here the use of the M05 templates
implies the derivation of lower ages and stellar masses. At older ages
the M05-templates are instead {\it bluer}.
\section{Galaxy data and SED Fitting}
For the present test we use the sample of seven galaxies in the Hubble
Ultra Deep Field (HUDF) which were singled out by the BzK criterion
for $z>1.4$ passively -evolving galaxies \citep{dadetal04}. These
galaxies show early-type morphologies on HUDF images and have
spectroscopic redshifts between 1.4 and 2.5 \citep[][hereafter
D05]{dadetal05}. Redshifts were determined from the Mg+Fe absorptions
at rest-frame $\lambda \sim 2600-2800$ \AA\ producing a characteristic
shape dubbed $Mg_{UV}$ feature by D05, on the HST+ACS grism spectra
acquired in the framework of the Grism ACS Program for Extra-galactic
Science project \citep[GRAPES,][]{piretal04}.  The presence of the
$Mg_{UV}$ feature imply that the UV/optical spectra of these galaxies
are consistent with being dominated by A- or F-type stars.  Therefore
these objects are not dominated by OB type stars like typical
star-forming galaxies and are predominantly passive. We choose these
particular sample for this study because we might expect that the
typical ages of stars in these galaxies are close to the critical
0.2-2~Gyr range where the different evolutionary synthesis models
differ by the largest amount. This is therefore an ideal sample of
high redshift galaxies to search for direct evidence of the effects
from TP-AGB stars.

Three of the seven galaxies were assigned a lower quality ("B" class)
redshift by D05, partly for the non perfect agreement with the
photometric redshifts. This is now improved in most cases with the
addition of the IRAC photometry to the SEDs and the M05 models (see
Table~2 and Figure~9).

Stellar population properties for these galaxies were derived in D05
by fitting their SEDs (ten bands from $B$ to $K$) to BC03 templates
spanning a wide range of parameters. Reddening was restricted to be
$E(B-V)<0.2$ and the Calzetti's law \citep{caletal00} was adopted. In
summary, the galaxies were found to be $\sim$ 1 Gyr old, to be in
passive evolution since at least $\sim 0.4$ Gyr, to have de Vauculeurs
light profiles and large stellar masses ($\gapprox
10^{11}~\msun$). They were therefore interpreted as progenitors of
massive ellipticals. The physical sizes of some of the objects ($\sim
0.7$ kpc) were found to be significantly smaller than objects with
similar masses in the local universe, which remained as an open issue
in D05.
\begin{figure}
\epsscale{1}
\plotone{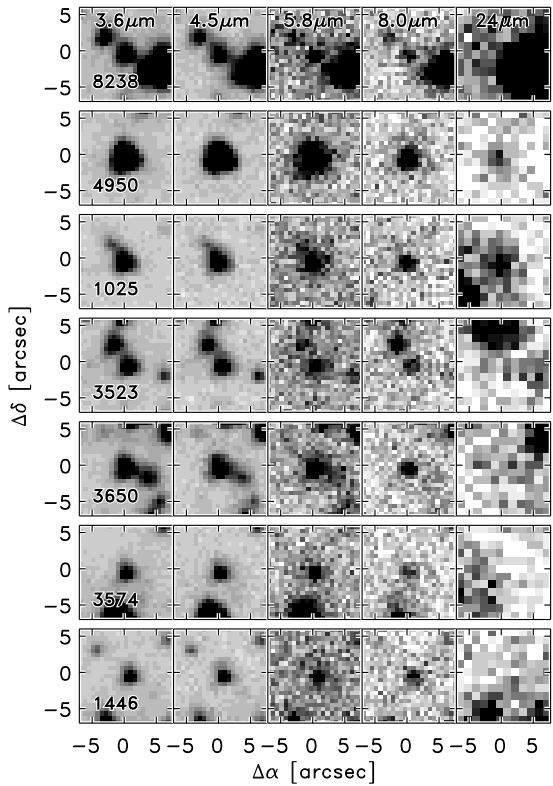}
\caption{The Spitzer IRAC and MIPS images of the seven HUDF galaxies. 
See Fig.~8 of D05 for HST ACS and NICMOS imaging. \label{images}}
\end{figure}
In the following we take advantage of the Spitzer IRAC (and MIPS)
imaging obtained over the Ultra Deed Field (UDF) as a part of the
Great Observatories Origins Deep Survey (GOODS) survey (Dickinson et
al., in preparation and Chary et al., in preparation,
respectively). All the seven galaxies were detected with high
signal-to-noise in the four IRAC bands (see Figure~\ref{images}). IRAC
photometry was obtained using $4''$ aperture magnitudes, corrected to
total using aperture corrections for unresolved sources.  The optical
to near-IR SEDs of D05 were based on Sextractor MAG\_AUTO
apertures. We measured large $6''$ diameter magnitudes in the $K$-band
to account for the flux lost by MAG\_AUTO, to perform the match with
the IRAC magnitudes. In order to account for the uncertainties in this
matching process, and in the IRAC aperture corrections, we added a
term of 0.1 mags in quadrature to the errors of the IRAC magnitudes.

\begin{figure}
\epsscale{1.}
\plotone{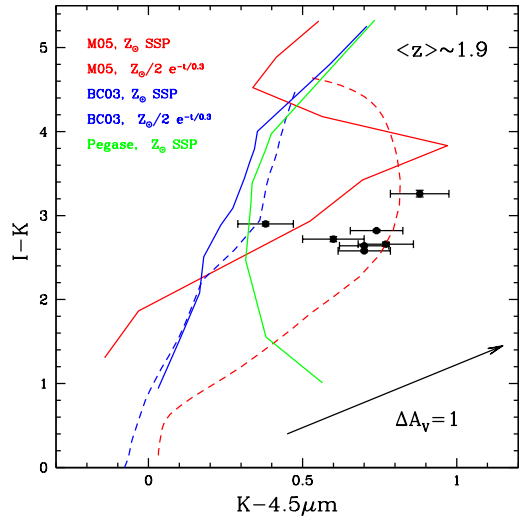}
\caption{Observed $I-K$ vs. $K-4.5\mu m$ colour-colour diagram for the
galaxies of our sample. Also shown are stellar population models from
different libraries (M05, BC03 and P\'egase) and for different star
formation histories, placed at the illustrative redshift of 1.9, with
ages ranging between 0.1 and 3.5 Gyr.\label{colcol}. The vector
indicates a reddening of $A_{\rm V}$= 1 mag for a Calzetti et al.~(2000) law \citep[from][]{labetal05}.}
\end{figure}
Figure~\ref{colcol} shows the observed $I-K$ vs. $K-4.5\mu m$
colour-colour diagram for our sample. Overplotted are some stellar
population models from the M05, BC03 and Pegase libraries, with ages
between 0.1 and 3.5 Gyr, in the observed frame at redshift 1.9. Simple
Stellar Populations (SSPs) with solar metallicity are indicated by
solid lines, while dashed lines display exponentially declining models
with e-folding time of 0.3 Gyr (for the P\'egase models, only the SSP
is shown). As in analogous plots by \citet{labetal05} and
\citet{papetal06}, the very red $K-4.5\mu m$ colours displayed by the
galaxies are not reached by the BC03 (or P\'egase) templates, unless
one adds strong dust reddening to the models. The M05 models match
instead the observed galaxy colours at the ages corresponding to the
AGB phase-transition.
The present database (see the Appendix for the photometric data)
allows the sampling of the galaxy spectral energy distributions up to
the rest-frame $K$-band. The whole wavelength range from the
rest-frame $UV$ to $K$ will be analysed in the next Section.
\subsection{SED fitting}
We use an adapted version of the code {\it Hyper-Z} \citep{boletal00},
kindly made available to us by M. Bolzonella, which performs SED
fitting at fixed spectroscopic
redshift\setcounter{footnote}{0}\footnote{For the fitting procedure we
use photometric errors of 0.05 if the formal error is smaller than
that, to account for systematics in photometry and color
matching}. The fitting procedure is based on maximum-likelihood
algorithms and the goodness of the fit is quantified via \chiquad
statistics (see the {\it Hyper-Z} manual for details). The code
computes the \chiquad for a certain number of templates, which differ
for star formation (SF) histories, metallicities and ages, and finds
the best-fitting template among them. It is important to note that the
code does not interpolate on the template grids. This implies that the
template set must be densely populated.

We have constructed templates that cover a wide range of stellar
population parameters, namely: i) metallicities from $\zsun/5$ to 2
$\zsun$; ii) ages from 10$^6$ yr to 5 Gyr (the maximum possible age
for the lowest redshift galaxy)\footnote{The spacing in age during the
runs is fixed by the {\it Hyper-Z} code, see the manual for details.};
iii) SF histories as: (a) instantaneous and chemically homogeneous
bursts (i.e. Simple Stellar Populations, SSPs); (b)
exponentially-declining modes (SFR $\propto e^{-t/ \tau}$, or
$\tau$-models, with $\tau=0.1,0.3,1$ Gyr); (c) {\it truncated} models,
where SF is constant for a finite time interval $t_{\rm trunc}$ and
zero thereafter ($t_{\rm trunc}=0.1,0.3,1,2$ Gyr); constant star
formation
\footnote{The models are available at
www-astro.physics.ox.ac.uk/$\sim\,$maraston}. These composite
templates (with prolonged SF histories) were constructed using the
BC03 software package, for consistency with D05 and in order to be
able to run the {\it Hyper-Z} code. As in D05, we determined 95\%
confidence ranges for ages and masses following \citet{avn76},
marginalizing over E(B-V), metallicity and SF history. This implies
that the determination of the latter parameters is formally less
constrained and the resulting values should be regarded as indicative.

To allow a direct comparison with other works, D05 in particular, the
templates refer to a straight Salpeter IMF down to 0.1
$M_{\odot}$. However, several pieces of evidence, from the redshift
evolution of the fundamental plane to the $M/L$ of local Es to the
metal content of the intergalactic medium strongly favour a
Salpeter-like IMF down to 1 $M_{\odot}$ but flatter below as required
to agree with the dynamically estimated $M/L$ ratios of early-type
galaxies \citep{ren05}. The actual shape of the IMF does not affect
appreciably the SED since at any age most of the light is produced by
stars in a narrow mass interval around the turnoff mass. However, it
impacts on the {\it amount} of light emitted per unit mass turned into
stars, thereby affecting the stellar $M^{*}/L$ ratio.  For example,
when using the IMF of \citet{kro01} the $M/L$ ratio is a factor $\sim
1.5$ smaller than in the straight-Salpeter case (M05).

For the objects classified of 'Class A redshift' by D05 the fitting
procedure included also the strength of the \mguv\ absorption feature,
as defined in D05 and measured on the GRAPES spectra. The difference
between observed and model \mguv\ strength was included in the
calculation of the \chiquadr\ when searching for the best-fit model,
with the same weight as the photometric data.

Finally, the entire SED fitting was repeated with BC03-based templates
spanning the same stellar population parameters as those used for the
M05 models.

The output of the procedure includes: the age $t$, i.e., the time since
the start of SF; the metallicity $[Z/H]$; the star-formation history
SFR; the reddening $E(B-V)$; and the stellar mass $M^{*}$.

The stellar mass $M^{*}$ is evaluated by comparing the observed and
the synthetic SED, which provides the mass that went into stars by the
age of the galaxy. This approach was retain in order to allow a
comparison with D05. However, such mass overestimates the true stellar
mass, as stars die leaving remnants whose mass is smaller than the
initial one. Adopting the prescriptions by \citet{rencio93} for the
initial mass-remnant mass relation, M98 evaluated that the stellar
mass of a SSP is reduced by $\sim 30\%$ 15 Gyr after formation, most
of which happens within the first few Gyrs. For given mass turned into
stars, such reduction is smaller in stellar populations with extended
star formation histories, due to a relatively higher fraction of
living stars.
\begin{figure}
\epsscale{1.}
\plotone{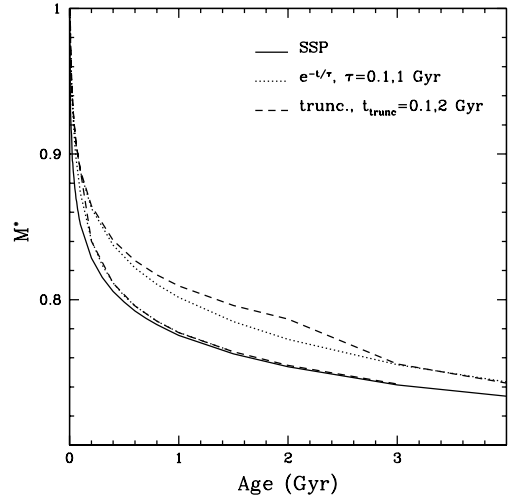}
\caption{The actual stellar mass $M^{*}$, normalised to an initial
value of $1~\msun$, of stellar populations with different star
formation histories.\label{remnants}}
\end{figure}
The actual $M^{*}(t)$ for some illustrative cases of star-formation
histories is given in Figure~\ref{remnants}. Tables 1 and 2 give the
stellar mass prior of such reduction, thus enabling a direct
comparison with other results. Table~2 also reports the mass decrement
(in percent) that corresponds to the given SFH and
age\footnote{Tabular values of $M^{*}$ for several star formation
histories can be found at~www-astro.physics.ox.ac.uk/$\sim\,$maraston.}. This mass decrement
should be applied to the masses listed in column 12.
\section{Results}
\subsection{Assuming no reddening, $E(B-V)=0$}
We first discuss the solutions that are obtained in the assumption
of zero reddening.
This allows us to single out the effect of stellar population modeling, 
thus highlighting 
some interesting differences between the two sets of models.
\begin{figure*}
\epsscale{1.2}
\plotone{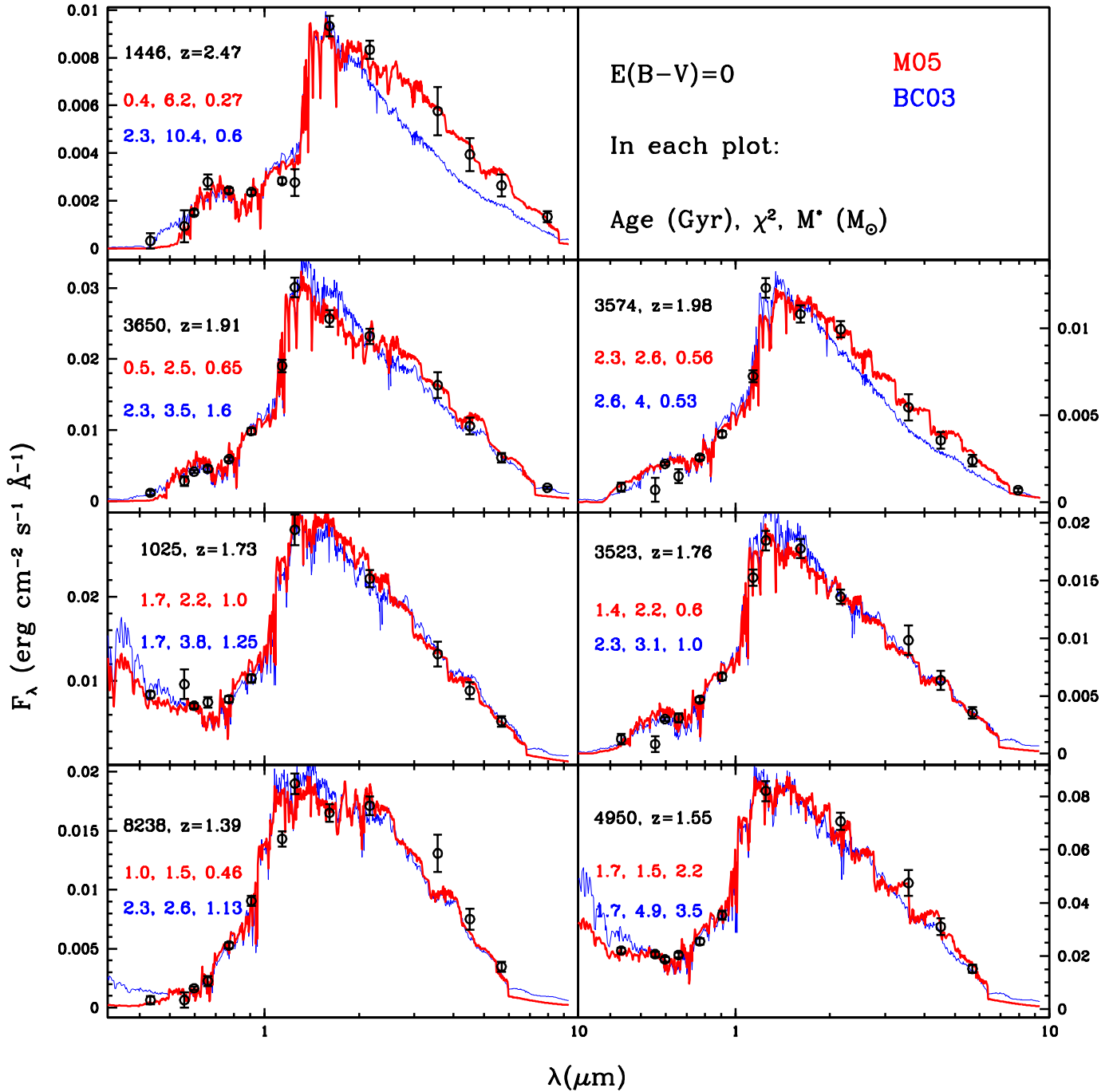}
\caption{The observed spectral energy distributions (filled symbols
with errorbars) and the best fit solutions obtained with the M05 and
the BC03 templates (red and blue lines, respectively). The parameters
of the fits are indicated on the figure and given in Table 1. No
reddening was adopted (i.e. E(B-V)=0).\label{fitnored}}
\end{figure*}
The corresponding best-fit solutions (lowest \chiquadr) for using the
BC03 (blue lines) and the M05 (red line) models are shown in
Figure~\ref{fitnored}. The observed SEDs are shown as solid symbols
with errorbars, and information on the individual galaxies and best
fit SFHs is given in Table~1.  Overall, the M05 models allow for
better fits (lower \chiquadr) for all galaxies, in the rest-frame
near-IR as well as in the blue (e.g., the objects at $z$=1.39, 1.55
and 2.47). Given the ages required to fit the observed SEDs, it is
likely that the different recipes for the TP-AGB phase are primarily
responsible for such effect. For the objects at $z=1.91$ and $z=2.47$
none of the BC03-based templates was able to match the near-IR fluxes.
In the case of the $z=2.47$ galaxy not even an age as old as 2.3 Gyr
is sufficient to obtain enough near-IR flux. Note also that the SSP
solution gives a fairly good fit with M05 models (which are TP-AGB
dominated at this age), whereas an extended SF history (constant until
$t_{\rm trunc}=2~\rm Gyr$) was required when using the BC03 models. In
this case the M05 single-burst models supply both the high optical and
the high near-IR fluxes, which come from the warm turnoff and cool
TP-AGB stars, respectively. The stellar masses $M^{*}$ are typically
smaller for the M05 models, being on average $\sim 60\%$ of those
derived with the BC03 models.  In part this is a consequence of the
lower ages, with the average age of M05 models being also $\sim 60\%$
of that derived for the BC03 models. The differences are not as high
as they would be in case of single bursts, as the more extended SFHs
required by the BC03 models have lower $M^{*}/L$ ratios.

The most relevant difference in the results from the two sets of
models is about the implied formation redshifts (see Table 1). Using BC03
models the stellar populations form at higher redshifts (sometime at
much higher redshifts, see e.g., the object at $z=2.47$) and, since
the galaxies are also more massive compared to the case when using M05
models, an appreciably faster mass growth of galaxies is implied.

\begin{figure*}
\epsscale{1.2}
\plotone{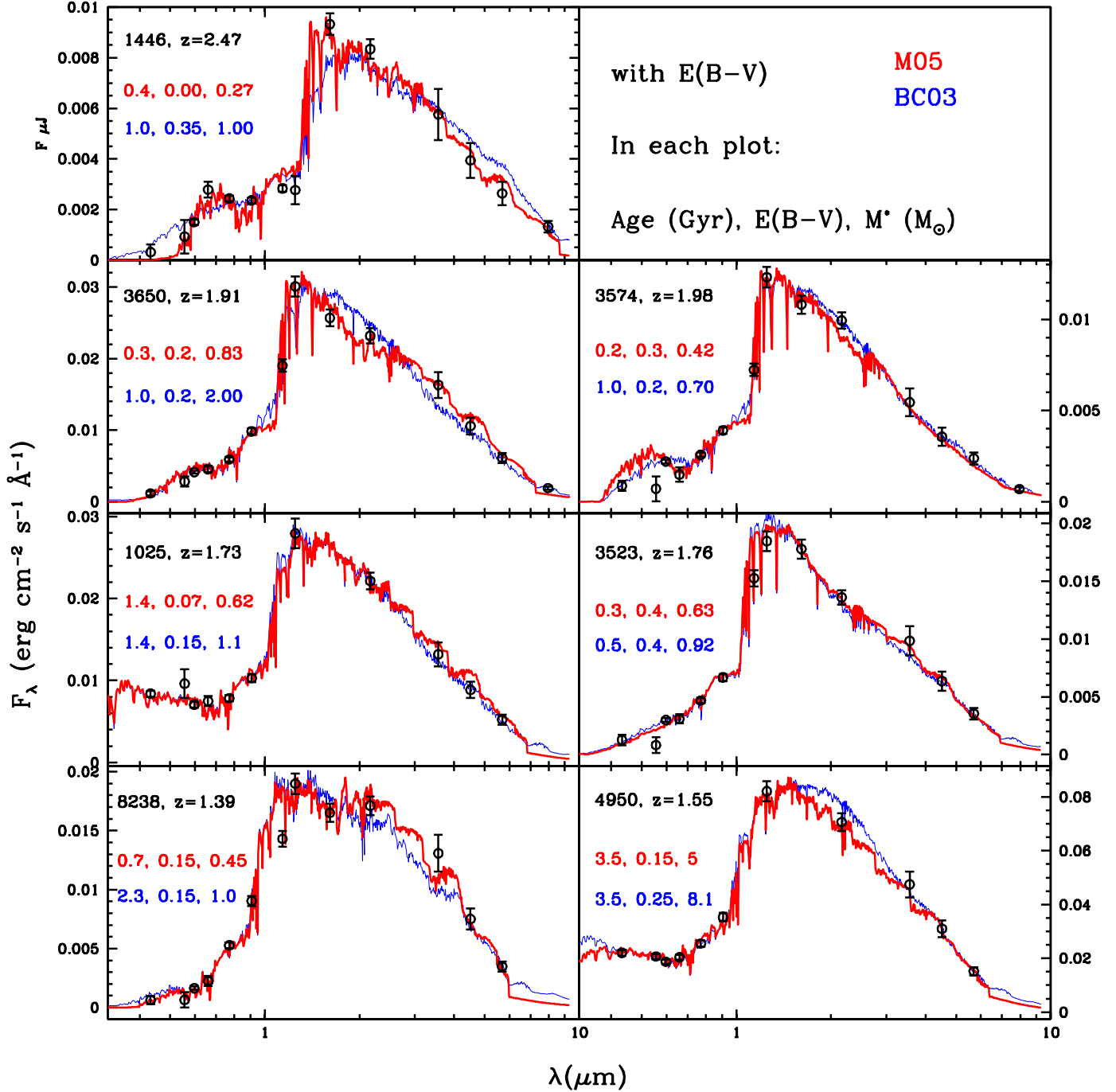}
\caption{Same as Figure~\ref{fitnored} but for $E(B-V)\neq 0$. The
parameters of the best-fits are provided in Table 2. 
\label{fitred}}
\end{figure*}

\subsection{$E(B-V)$ as a free parameter}

The best-fit solutions when also reddening is treated as a free
parameter are displayed in Figure~\ref{fitred}, and the parameters of
the best-fits are given in Table 2. 
The reddening was allowed to vary among the options offered in the
{\it Hyper-Z} package, namely, Milky Way, SMC, LMC and the so-called
Calzetti's law for local starbursts (references in Table 2).  All
these prescriptions refer to a dust-screen configuration. 
For this 'reddening' case, in
order to exploit all the available information and to control the
degeneracies introduced by the larger number of free parameters, we
have additionally included the strength of the \mguv\ feature in the
computation of the \chiquad. This was done only for galaxies with
Class A redshift, namely for objects 4950, 1025, 3650, 3574, because
for the Class B objects the \mguv\ is a less robust measurement (D05).

By allowing the best-fit procedure to treat $E(B-V)$ as an additional
free parameter, and allowing it to choose among four different
reddening laws, it is no surprise that solutions with appreciably
better \chiquad are found. However, this does not ensure that such
solutions are more likely than those assuming $E(B-V)=0$. It is also
no surprise that some best-fits require the Calzetti's law, whereas
others prefer the SMC law. In just one case the LMC law was
preferred, while in no cases the MW law was chosen. It is worth noting
that the best-fit for the $z=2.47$-object exhibiting the strongest
TP-AGB contribution is reddening-independent for the M05 models.

In general, the inclusion of reddening reduces the ages of the
best-fits obtained with the BC03 models, making them closer to those
obtained with the M05 models. However, reddening
does not help to improve substantially the fits to the IRAC bands,
like in the case of objects 1446 and 3650. TP-AGB effects cannot be
traded with dust or metallicity effects, which helps reducing the
degeneracies in the age determination.
Detailed comments on the fits for individual galaxies  are given 
in the Appendix.\\
\begin{figure}
\epsscale{1}
\plotone{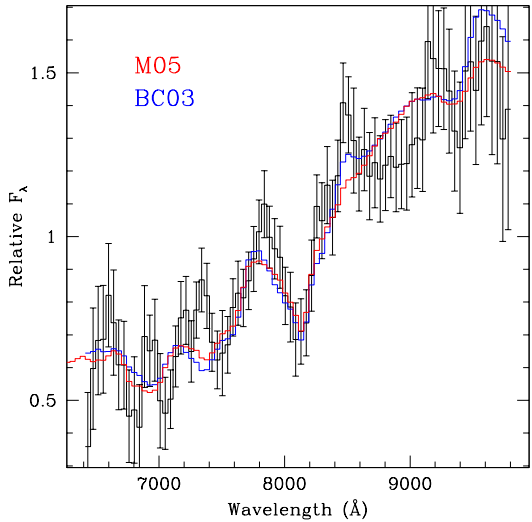}
\caption{The GRAPES HST+ACS spectrum of galaxy 3650 at $z\sim 1.91$
  (from \citet{dadetal05}), with superimposed the best-fits obtained
  with the BC03 and the M05 models (blue and red lines,
  respectively, smoothed at the same spectral resolution of the
  observed spectrum). \label{grapes}}
\end{figure}

Figure \ref{grapes} compares the GRAPES HST+ACS spectrum of object
3650 at $z \sim 1.91$, with the best-fit solutions corresponding to
the BC03 and M05 models (see Table~2). We use this plot for
illustrating the effect of the age/metallicity degeneracy that affects
the optical, but is alleviated in the near-IR due to the distinctive
spectral features of TP-AGB stars. The best-fit solutions recover
similarly well the strength of the \mguv\ line and the spectral shape
around it, in spite of having different population parameters, namely
young and metal-rich for the M05 models and older and metal-poor for
the BC03 models. However, the BC03-based solution gives an appreciably
poorer fit to the near-IR SED (cf. Figure~\ref{fitred}), showing how
the inclusion of the near-IR fluxes can help breaking the
age/metallicity degeneracy.

However in most cases spectral absorptions and photometric data on a
wide spectral range are not available simultaneously. How realistic
are then the best-fits obtained from the sole photometry? To get
insight on this issue, we have repeated the fitting procedure without
including the \mguv\ line when searching the best-fit model.  We have
then checked how well these 'photometric' best-fit models recover the
observed \mguv\ .
\begin{figure}
\epsscale{1.25}
\plotone{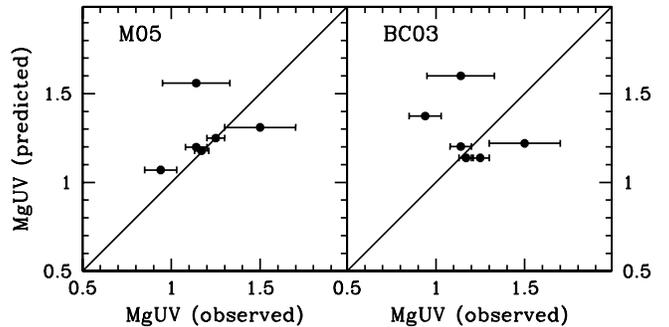}
\caption{Comparison between the measured $Mg_{\rm UV}$ indices(from
D05) and the values predicted by the best-fit models obtained with the
M05 and the BC03 templates using only the photometric
SEDs.\label{mguvnochi}}
\end{figure}
This is shown in Figure~\ref{mguvnochi}, in which the \mguv\ strengths
predicted by the best-fit `photometric' models are compared to the
observed ones. Although based on a very small number of objects, the
predicted values from M05 models are somewhat more accurate than those
from the BC03 models (variances are 0.04 and 0.08, respectively).

\begin{figure}
\epsscale{1.25}
\plotone{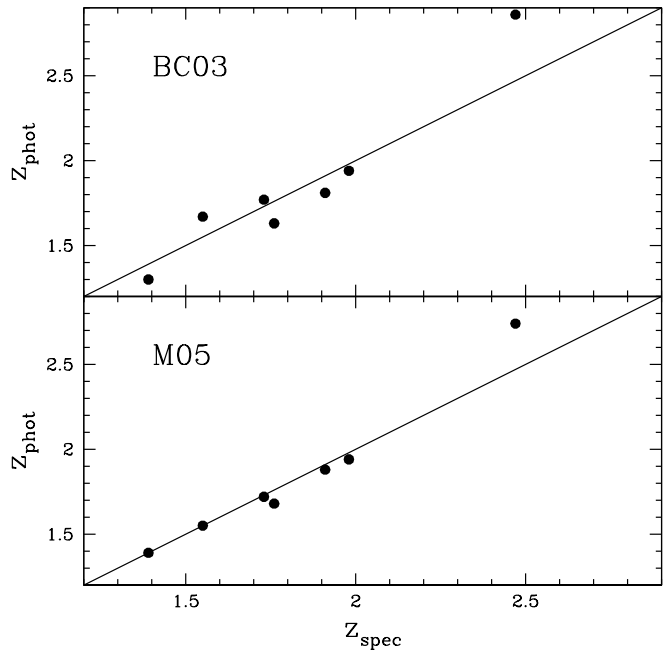}
\caption{Comparison between the spectroscopic redshifts and the
photometric redshifts obtained with the BC03 and M05 models. The IRAC
magnitudes have been included in the derivation of the photometric
redshifts.\label{photoz}}
\end{figure}
As a final exercise, we have checked the effect on photometric
redshifts derived using one or the other set of population models, and
of the inclusion of the IRAC fluxes. The results are given in Table 2
and shown in Figure~\ref{photoz}. The photometric redshifts obtained
with the M05 templates are in better agreement with spectroscopic
redshifts than those obtained with the BC03 models. The photometric
redshifts obtained with the BC03 models and the IRAC data appear to be
more discrepant than those in which only magnitudes up to $K$ were
used (cf. D05). Also this effect likely originates from the different
TP-AGB recipes.
\subsection{Stellar Masses}
\begin{figure*}
\epsscale{1}
\plotone{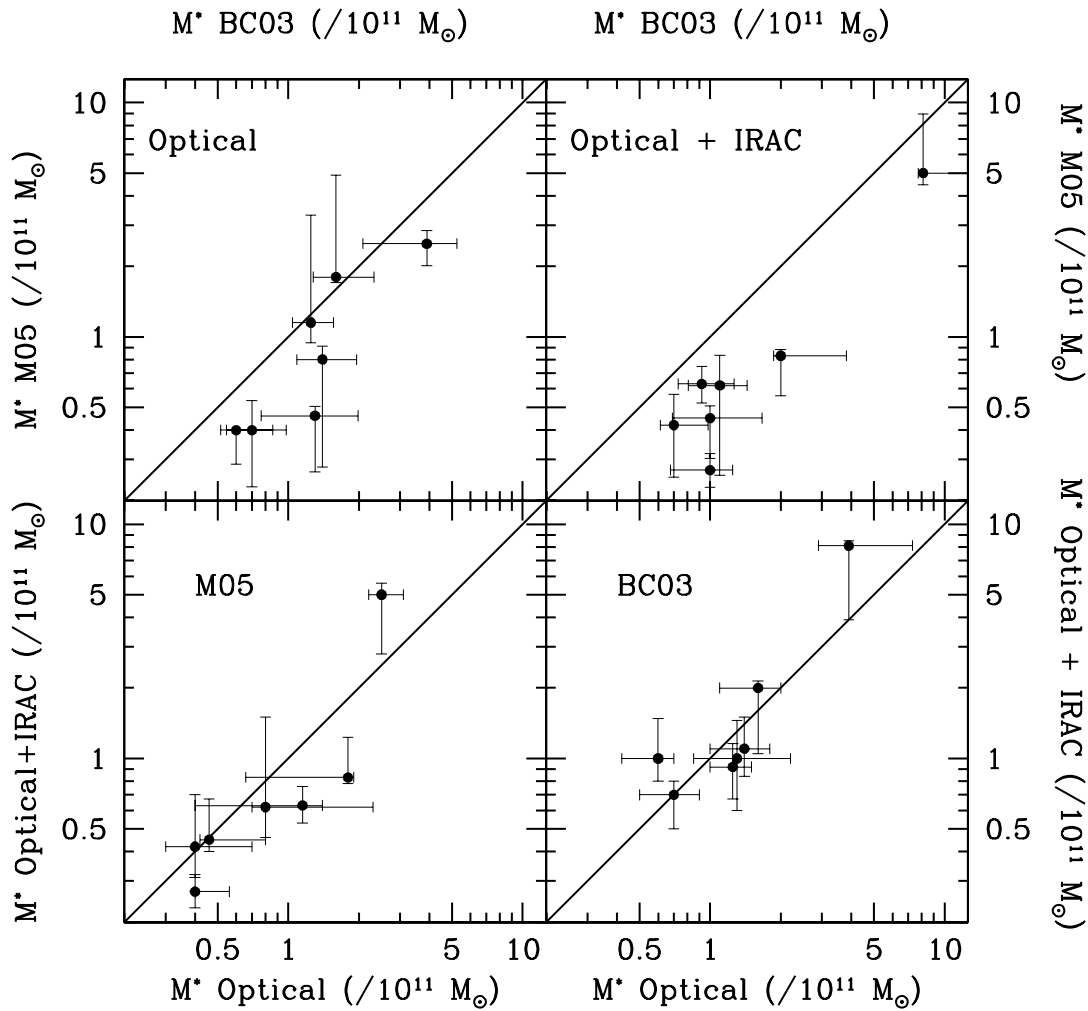}
\caption{{\it Upper panels}. Effect of the stellar population
templates on the stellar masses $M^{*}$, using only optical bands (to
observed $K$, left-hand panel) and optical plus IRAC (right-hand
panel). {\it Lower panels}. Effect of the wavelength range for given
set of templates, M05 (left-hand panel) and BC03 (right-hand
panel). The \mstar plotted here were not decreased by the stellar
mass-losses for consistency with D05. \label{mass}}
\end{figure*}

The different star-formation histories derived from the different
models have an impact on the derived stellar masses, which is
summarised in Figure~\ref{mass}. The upper panels display the effects
of the models (i.e., M05 vs BC03) on the derived stellar mass of
galaxies, when using only the optical and near-IR bands (left-hand
panel) and when also the IRAC 3.5--8 $\mu$m data are used (right-hand
panel). The masses predicted by the M05 templates are lower, which is
a consequence of the lower ages. The effect is appreciably more
pronounced when the IRAC bands are included, as the TP-AGB phase (the
most discrepant ingredient between the two models) has its largest
contribution in the rest-frame near-IR. However, lower masses can also
be derived when the observed-frame $K$ is the reddest band used in the
best fit. In fact, for the objects at $z=1.39.1.55,1.73$ the $K$ band
samples the rest-frame $I$, which is already affected by the TP-AGB
(M05, Figure 18). In addition, the later onset of the RGB phase in the
overshooting tracks adopted by BC03 helps to make the BC03 templates
fainter at ages around 0.5 Gyr. The corresponding $M/L$ ratio is
higher and a higher \mstar is derived (see Section 2).  It is
embarrassing to realize to which extent the recipes for the convective
overshooting impact on the stellar masses that are derived for
high-$z$ galaxies, and therefore on the implied galaxy formation
scenarios.

The lower panels in Figure \ref{mass} show the effects (for given
stellar population models) of the wavelength range included in the
fitting procedure: only up to the $K$ band (on the x axis) and with
the IRAC bands (on the y axis). The results based on BC03 are not
appreciably affected by the inclusion of the IRAC bands \citep[as
recently pointed out by][]{shaetal05}. Instead, for the M05-based
solutions, three galaxies are found to have appreciably lower masses
when the IRAC fluxes are considered, again the likely effect of the
inclusion of the TP-AGB contribution.

\begin{figure}
\epsscale{1}
\plotone{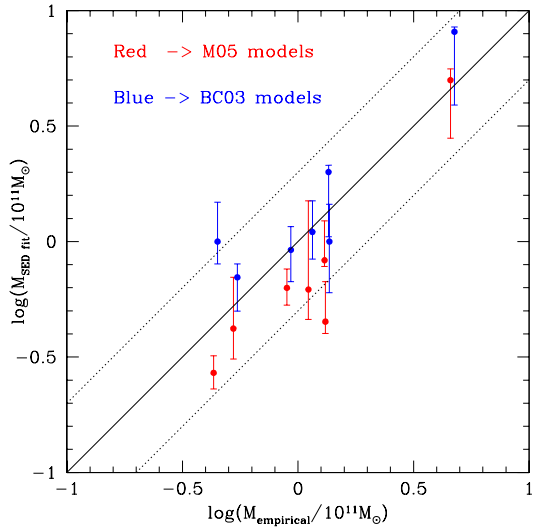}
\caption{Comparison of the stellar masses obtained in this paper
(Table~2, Column 12) with the calibration proposed by
\citet{dadetal04}. Dotted lines show factor 2 variations around the
average calibration. \label{bzk}}
\end{figure}

Finally, Figure \ref{bzk} compares the masses obtained in this paper
(including the IRAC data) with the calibration for $z\sim2$
$BzK$-selected galaxies proposed by \citet{dadetal04}. This
calibration was obtained using the observed $K$-band magnitude and
$z-K$ colour only, along with the BC03 templates. The figure shows
that the calibration is well consistent with the BC03-based masses
derived in this paper, which is no surprise. The M05 models suggest
instead a calibration lower by 0.2~dex, on average.
\subsection{MIPS detection and ongoing SFR}
Only two galaxies are significantly detected with Spitzer+MIPS at
24$\mu m$ (objects 1025 and 4950, Fig.~\ref{images}).  The MIPS 24$\mu
m$ fluxes of about 25--30$\mu$Jy convert to star-formation rates of
about 20-30 $M_\odot$yr$^{-1}$ at the redshifts of the sources, using
the models from \citet{charelb01}. The two MIPS detections are in fact
the galaxies with the bluer UV spectra (see D05) and more extended
star formation histories (see Table~2). The MIPS derived SFRs are
comparable, given the uncertainties, with the residual ongoing SFRs
(of $\sim 10-20 \msun/$yr) that are obtained through the SED modeling
of these 2 galaxies (see also D05).  For the rest of the galaxies the
3~$\sigma$ limits on the 24$\mu m$ flux imply SFR below 10
$M_\odot$yr$^{-1}$. We recall that typical $z=2$ galaxies with similar
K-band fluxes to the sample studied here have much stronger 24$\mu m$
emission, of order of 120$\mu$Jy \citep{dadetal05b}. The low MIPS
fluxes from these sources underline once again, in agreement with
their spectral features, that these objects are predominantly passive.
\subsection{The role of AGNs}
In obtaining the best-fitting models we have assumed that the observed
SEDs are the results of just stars and dust. The possible presence of
an AGN would obviously affect stellar population modeling in general.
However, differences between the physical properties of galaxies
inferred with the M05 or BC03 models remain, as these models have a
different ratio of optical to near-IR flux. Therefore, even if at
play, AGNs would not alter the sense of this investigation.

More specifically, AGNs probably play a minor role on our analysis.
First, no clear AGN features are observed in the spectra of these
objects. Also, the shape of the SED over IRAC is clearly stellar with
a secure identification of the peak emission at 1.6 $\mu m$. Two of
the galaxies (1025 and 1446) have $X$-ray detections, but as discussed
in D05 these are extremely hard and absorbed. Coupled to the evidence
of the faintness of the galaxies at 24 $\mu m$, we conclude that the
AGN should not influence appreciably the photometry from the UV to the
near-IR rest-frame. For example, if we adopt the SED shapes of NGC1068
or MRK231 and normalise them to the 24$\mu m$ photometry of the
sources, the contribution of the AGN to the optical to near-IR light
of the galaxies would be entirely negligible. Similarly, from the
narrow $K$-band to $X$-ray flux correlation of AGN shown by
\citet{bruetal05}, we conclude that possible AGN in our sources are
expected to provide a negligible contribution to the $K$-band light of
our galaxies.

Finally, the overall good agreement between predicted and observed
\mguv\/ line-strengths (Figure~7) and the good match between
spectroscopic and photometric redshifts (Figure~8) suggest that a
significant AGN contribution is highly unlikely.

\section{Discussion}
The aim of this paper is to explore how much the star formation
histories of high-$z$ galaxies, that are derived by fitting observed
to synthetic spectral energy distributions depend on the adopted
stellar population models. In particular, we focused on the TP-AGB
evolutionary phase, because TP-AGB stars are the dominant bolometric
and near-IR contributors in stellar populations with ages $\sim$ 1 Gyr
(M98; M05). Therefore, such phase must be of primary importance at
high-$z$, when the universe itself was just a few Gyrs old
\citep{mar04,mar05,ren92}, and especially in the interpretation of
{\it Spitzer} data for high-$z$ objects. The recipes for the TP-AGB
phase are found to differ significantly among the various models,
which has motivated us to check the robustness of the results obtained
for high-$z$ galaxies.

To this aim we have analysed a sample of mostly passively evolving
high-$z$ ($1.4 \lapprox z \lapprox 2.5$) galaxies with optical and
{\it Spitzer}/IRAC photometry, modeling the whole spectral energy
distribution from the rest-frame $B$ to $K$. The analysis was
performed with two libraries of stellar models: the M05 models in
which the TP-AGB phase is calibrated on Magellanic Cloud globular
clusters, and the BC03 models in which, like in most other models in
the literature (e.g. Pegase, Starburst99) the energy contribution from
the TP-AGB phase is significantly lower than in the Maraston
models. Furthermore, the two models adopt stellar evolutionary tracks,
in which the onset of the Red Giant Branch is predicted at different
ages, which also impact on the time evolution of the synthetic
spectral energy distribution.  The differences between stellar
population models are widely discussed in M05 and recalled in Section
2.

Besides the 14 photometric bands spanning from rest-frame $B$ to $K$,
the procedure introduced in this paper also includes the strength of a
near-UV absorption feature (the \mguv\ at 2800 \AA, see D05) for
searching the best-fit galaxy models. In this way the effect of the
age/metallicity degeneracy is substantially reduced. We have also
checked that even without using the spectroscopic information, but
only with the photometric SED, the best-fits obtained with the M05
models recover well the strength of the \mguv\ feature, hence provide
consistent models for distant galaxies.

We then show to which extent the derived star-formation histories of
the sample galaxies are model dependent. The M05 models typically
match the rest-frame near-IR and optical very well, without invoking
strong reddening and without violating the constraint set by the age
of the universe. The BC03-based solutions exhibit in some cases
near-IR fluxes that are significantly lower than the data, which
requires the addition of old stellar populations or dust reddening or
a combination of the two. This is reminiscent of recent results on
high-$z$ galaxies in which Spitzer/IRAC data are interpreted with the
BC models \citep[e.g.,][]{yanetal04} or the Pegase models
\citep[e.g.,][]{viletal06}. The young solutions required to match the
blue side of the SED could not match the rest-frame near-IR, which led
the authors to assume an underlying old population to be responsible
for the near-IR fluxes. While a two-component model is not implausible
in itself, the proportions by mass of the young and old component (1
vs. 99 \%) as well as the age of the old component, comparable to the
age of the Universe, make the star formation history contrived.

The lower masses derived from the M05 models may help understanding
the origin of an apparent discrepancy noted by D05 and concerning the
relationship between the physical sizes and the masses of the galaxies
in the present sample. The galaxies were found to be too small for the
given masses in comparison to local objects. The lower masses derived
here already help in alleviating the discrepancy. In particular, the
$z=2.47$ object, with a stellar mass around $10^{10}~\msun$, is more
likely the precursor of a bulge rather than a massive elliptical. Its
rather small size ($r_{\rm e}\sim 0.8$ kpc, D05) supports this
conclusion, as it can be inferred from comparisons with local samples
\citep{benetal92}. Also its formation redshift and metallicity
compares well with similar estimates of bulges in the local universe
\citep{thodav06,sarjab05}. According to the BC03 models, three
galaxies have rather low metallicities ($\zsun/5$) and at the same
time high masses ($>10^{11}~\msun$). Such objects do not match local
relationships, even taking metallicity gradients into account. Only
one such object is found with the M05 models.

Finally, we find that the stellar population templates impact on
photometric redshift determination. The use of IRAC bands and M05
templates give photometric redshifts that are in excellent agreement
with the spectroscopic ones. Again, we ascribe this effect to the
prescriptions for the TP-AGB phase.

In this work we have focused on high-$z$ galaxies, where the stellar
populations are young therefore TP-AGB stars must exist. Our spectral
fitting represents the first direct evidence for such stars in the
early universe. Going to even higher redshifts might reveal further
interesting aspects of galaxy evolution, a scope we are pursuing in
the future.  However, TP-AGB stars may matter at any redshift
depending on the specific star formation history. For example,
\citet{vdw05} measured the evolution of the rest-frame $K$-band
Fundamental Plane from $z \sim 1$ to the present, by using IRAC
imaging of a sample of early-type galaxies. They find that the
$K$-band luminosities of the low-mass objects are too large (at given
$B$ luminosity), or the $B-K$ colours too red, with respect to the
prediction of the BC03 or \citet{vazetal96} models. The M05 models
were found to provide a better match the data. This effect, which is
correctly attributed to the treatment of the TP-AGB phase, implies
that those low-mass early-type galaxies experienced a burst of star
formation at redshift close to one.

In conclusion, the recipe for the stellar population modeling plays a
central part in the derivation of the star formation histories of
high-$z$ galaxies, which should be kept in mind when the derived
parameters are used to constrain galaxy formation.
\acknowledgments
We thank Micol Bolzonella for assistance with the $Hyper-Z$ code. CM
ackwnoledges discussions with Reinhard Genzel. CM is a Marie Curie
fellow and holds grant MEIF-CT-2005-011566 of the Training and
Mobility of Researchers programme financed by the European
Community. ED acknowledges NASA support through the Spitzer Fellowship
Program, award 1268429. Some of the data used here are part of the
GOODS Spitzer Space Telescope Legacy Science Program, that is
supported by NASA through Contract Number 1224666 issued by the JPL,
Caltech, under NASA contract 1407.
%

%
\clearpage
\begin{deluxetable}{ccccccccc}
\tabletypesize{\scriptsize}
\tablecaption{Galaxy SED fitting results: no reddening}
\tablewidth{13cm}
\tablehead{
\colhead{ID\tablenotemark{a}} & \colhead{$z_{\rm spec}$} & mod & \colhead{t} & \colhead{$z_{\rm form}$} & \colhead{$\zh$} & \colhead{SFH} & \colhead{$\chi^{2}_r$} & \colhead{$M_{*}$\tablenotemark{b}} \\
 & & & Gyr & & & & $M_{\odot}/10^{11}$ 
}
\startdata
 8238 & 1.39 & M05 & 1.0 & 1.8 & 2 $\rm \zsun$ & $\rm e^{-t/0.1~Gyr}$ & 1.5 & ${0.46}^{+0.48}_{-0.06}$ \\
 &  & BC03 & 2.3 & 2.9 & 2.5 $\rm \zsun$ & $\rm e^{-t/0.3~Gyr}$ & 2.6 & ${1.13}^{+1.6}_{-0.3}$ \\
 4950 & 1.55 & M05 & 1.7 & 2.7 & $\rm \zsun/2$ & $\rm e^{-t/0.3~Gyr}$ & 1.5 & ${2.20}^{+0.3}_{-0.4}$ \\
 &  & BC03 & 1.7 & 2.7 & 2.5 $\rm \zsun$ & $\rm e^{-t/0.3~Gyr}$ & 4.9 & ${3.50}^{+5.8}_{-0.6}$  \\
 1025 & 1.73 &  M05 & 1.7 & 3.2 & $\rm \zsun$ & $\rm e^{-t/0.3~Gyr}$ & 2.2 & ${1.00}^{+0.4}_{-0.3}$ \\
 &  & BC03 & 1.7 & 3.2 & 2.5 $\rm \zsun$ & $\rm e^{-t/0.3~Gyr}$ & 3.8 & ${1.25}^{+1.35}_{-0.2}$  \\
 3523 & 1.76 & M05 & 1.4 & 2.9 & $\rm \zsun$ & $\ttru=1.0$ Gyr & 2.2 & ${0.60}^{0.5}_{1.06}$  \\
 &  & BC03 & 2.3 & 4.6 & 2.5 $\rm \zsun$ & $\ttru=2.0$ Gyr & 3.1 & ${1.00}^{+0.4}_{-0.3}$  \\
 3650 & 1.91 & M05 & 0.5 & 2.2 & 2 $\rm \zsun$ & $\ttru=0.1$ Gyr & 2.5 & ${0.65}^{+0.55}_{-0.01}$ \\
 &  & BC03 & 2.3 & 5.5 & 2.5 $\rm \zsun$ & $\ttru=2.0$ Gyr & 3.5 & ${1.6}^{+0.2}_{-0.5}$  \\
 3574 & 1.98 & M05 & 2.3 & 6 & $\rm \zsun/5$ & $\ttru=1.0$ Gyr & 2.6 & ${0.56}^{+0.14}_{-0.10}$ \\
 &  & BC03 & 2.6 & 8.2 & 0.4 $\rm \zsun$ & $\ttru=2.0$ Gyr & 4.0 & ${0.53}^{+0.13}_{-0.15}$ \\
 1446 & 2.47 & M05 & 0.4 & 2.8 & 2 $\rm \zsun$ & SSP & 6.2 & ${0.27}^{+0.10}_{-0.02}$ \\
 &  & BC03 & 2.3 & 15 & $\rm \zsun$ & $\ttru=2.0$ Gyr & 10.4 & ${0.60}^{+0.08}_{-0.18}$  \\
\enddata
\tablecomments{The variable $t$ is the temporal distance from the
beginning of star-formation. The formation redshift $z_{\rm form}$
refers to $t$, thereby indicating, in case of extended star formation
histories, the epoch at which the galaxy started to form. The last
IRAC filter was excluded for objects up to z=1.76, for which this
filter samples rest-frame wavelengths $>$ 2.5 microns, that is the
limit where the empirical TP-AGB star spectra are defined in the M05
models. The same filter configuration was used when the BC03 models
were employed.  We use the following conversion between AB and Vega
magnitudes in the $B$- and $K$-band: $M_{\rm B(Vega)}=M_{\rm
B(AB)}-(-0.089)$, $M_{\rm K(Vega)}=M_{\rm K(AB)}-(1.871)$ .}
\tablenotetext{a}{Galaxy ID as in Daddi et al. 2005}
\tablenotetext{b}{The mass is obtained from normalisation of the SED.
Ranges refer to the 95\% confidence level. Note that the total
magnitudes are defined from the values obtained from S\'ersic profile
fitting given in D05, Table~1.}
\end{deluxetable}
\begin{deluxetable}{lccccccccccccl}
\tabletypesize{\scriptsize}
\tablecaption{Galaxy SED fitting results}
\tablewidth{20cm}
\tablehead{
\colhead{ID\tablenotemark{a}} & \colhead{$z_{\rm spec}$} & \colhead{$z_{\rm phot}$} & mod & \colhead{t} & \colhead{$z_{\rm form}$} &\colhead{$\zh$} & \colhead{SFH} & \colhead{E(B-V)} & \colhead{$\chi^{2}_r$} & \colhead{$M_{*}$\tablenotemark{e}} & \colhead{$M_{*}$\tablenotemark{f}} & \colhead{$M_{*}$\tablenotemark{g}} & \colhead{$\% M$\tablenotemark{h}} \\
 & & & & Gyr & & & & & $M_{\odot}/10^{11}$ & $M_{\odot}/10^{11}$ & $M_{\odot}/10^{11}$ &  
}
\startdata
 8238 & 1.39 & 1.39 & M05 & 0.7 & 1.7 & 2 $\rm \zsun$ & $\ttru=0.3$ Gyr & 0.15\tablenotemark{b} & 1.1 & 0.25$-$0.47 & ${0.45}^{+0.22}_{-0.05}$ & 1$-$2.4 & 21 \\
 &  & 1.30 & BC03 & 2.3 & 2.9 & 2.5 $\rm \zsun$ & $\ttru=2.0$ Gyr & 0.15\tablenotemark{c} & 2. &  & ${1.0}^{+0.45}_{-0.40}$ & \\
 4950 & 1.55 & 1.55 & M05 & 3.5 & 7.5 & $\rm \zsun/2$ & $\rm e^{-t/1Gyr}$ & 0.15\tablenotemark{b} & 0.6 & 2.8$-$5.6 & ${5.00}^{+0.6}_{-2.2}$ & 2.9$-$7.3 & 26 \\
 &  & 1.67 & BC03 & 3.5 & 7.5 & $\rm \zsun/5$ & $\rm e^{-t/1Gyr}$ & 0.25\tablenotemark{c} & 0.94 &  & ${8.10}^{+0.4}_{-4.2}$ & \\
 1025 & 1.73 & 1.72 & M05 & 1.4 & 3.1 & $\rm \zsun$ & $\rm e^{-t/0.3Gyr}$ & 0.07\tablenotemark{b} & 1.1 & 0.3$-$0.4 & ${0.62}^{+0.88}_{-0.16}$ & 1$-$1.8 & 20 \\
 &  & 1.77 & BC03 & 1.4 & 3.1 & $\rm \zsun/2.5$ & $\rm e^{-t/0.3Gyr}$ & 0.15\tablenotemark{c}& 0.64 &  & ${1.1}^{+0.4}_{-0.26}$ & \\
 3523 & 1.76 & 1.68 & M05 & 0.3 & 1.9 & $\rm \zsun$/5 & $\ttru=0.1$ Gyr & 0.4\tablenotemark{b} & 1.2 & 0.4$-$0.6 & ${0.63}^{+0.13}_{-0.1}$ & 1$-$1.5 & 18 \\
 &  & 1.63 & BC03 & 0.5 & 2.1 & $\rm \zsun$ & $\rm e^{-t/0.1Gyr}$ & 0.4\tablenotemark{b} & 1.2 & & ${0.92}^{+0.24}_{-0.25}$ & \\
 3650 & 1.91 & 1.88 & M05 & 0.3 & 2.1 & 2 $\rm \zsun$ & SSP & 0.2\tablenotemark{b} & 1.3 & 0.33-0.85 & ${0.83}^{+0.40}_{-0.05}$ & 1.3$-$2 & 19 \\
 &  & 1.81 & BC03 & 1.0 & 2.7 & $\rm \zsun$/5 & SSP & 0.2\tablenotemark{c} & 1.9 & & ${2.0}^{+0.14}_{-0.95}$ & \\
 3574 & 1.98 & 1.94 & M05 & 0.2 & 2.1 & 2 $\rm \zsun$ & SSP & 0.3\tablenotemark{h}& 2.5 & 0.2$-$0.4 & ${0.42}^{+0.28}_{-0.11}$ & 0.5$-$0.9 & 18 \\
 &  & 1.94 & BC03 & 1.0 & 2.9 & $\rm \zsun/5$ & $\ttru=0.3$ Gyr & $0.2$\tablenotemark{c} & 2. & & ${0.7}^{+0.3}_{-0.2}$ & \\
 1446 & 2.47 & 2.74 & M05 & 0.4 & 2.8 & 2 $\rm \zsun$ & SSP & 0.00 & 3.1\tablenotemark{i} & 0.27$-$0.2 & $0.27^{+0.05}_{-0.04}$ & 0.7$-$1.1 & 20 \\
 &  & 2.86 & BC03 & 1.0 & 3.8 & $\rm \zsun$ & $\rm e^{-t/0.1Gyr}$ & 0.35\tablenotemark{c} & 4.7\tablenotemark{i} & & $1.0^{+0.48}_{-0.2}$ \\
\enddata
\tablecomments{The best-fit parameters and \chiquadr\ refer to the
model obtained using the spectroscopic redshift $z_{\rm spec}$. The
values $z_{\rm phot}$ are given for comparison. See Table~1 for other
notes.}  \tablenotetext{a}{Galaxy ID as in Daddi et al. 2005}
\tablenotetext{b}{SMC reddening law as in \citet{preetal84} and
\citet{bouetal85}} \tablenotetext{c}{Reddening law as in Calzetti et
al. (2000)} \tablenotetext{d}{LMC reddening law as in \citet{fit86}}
\tablenotetext{e}{Through $M^{*}/L_{\rm B_{\rm Vega}}$ and
$M^{*}/L_{\rm K_{\rm Vega}}$} \tablenotetext{f}{See note (b) to Table 1}
\tablenotetext{g}{From Daddi et al. 2005} \tablenotetext{h}{This
column gives the percentage of mass decrement due to stellar mass
loss. It should be applied to values in Column 11 (labelled {\em f})}
\tablenotetext{i}{For this object the \chiquad\ were computed without
the photometric data in the Nicmos $F110W$ filter. This point is off
but has a small errorbar, which drives the \chiquad\ to the misleading
values of 6.2 and 5.8, in case of the M05 and BC03, respectively.}
\end{deluxetable}
\appendix
Here we comment in more details on the best-fits obtained for
individual objects for the reddening case (cfr. Table~2).

{\bf 8238, z=1.39}. The \mguv\ line was not included in the
\chiquad\. The object is $\sim 0.7$ Gyr old, metal-rich and in
passive evolution since 0.4 Gyr. The reddening follows an SMC-like
law. The stellar mass is $0.45\cdot 10^{11}\msun$ with small
dispersion. For the BC03 models, the best solution has a worse
\chiquad, a significantly older age and a significantly longer
formation timescale. The stellar mass is a factor 2.2 larger and
consistent with what obtained in D05 from fitting the sole optical
bands.

It is interesting to note that the same, smaller stellar mass would
have been derived by fitting the M05 models up the sole observed-frame
$K$ (see Figure~\ref{mass}). This is due to two effects. First, at
redshift $\sim 1.4$ the observed $K$ samples into the rest-frame $I$,
which is already affected by the TP-AGB prescriptions (M05).  Second,
as pointed out in Section 2, another important difference between the
two sets of models exist, namely the energetics of the different input
stellar evolutionary tracks. The M05 models are based on tracks in
which the RGB phase starts developing at younger ages with respect to
the Padova tracks used in BC03. The delay in the latter is due to the
overshooting (see discussion in M05). The result is that the M05 SSP
models are brighter around 1 Gyr (cf. M05, Fig. 7), which implies a
lower stellar mass, an effect anticipated in M05. It should be noted
that the earlier onset of the RGB phase compares very well with the
observed RGB contribution in Magellanic Cloud globular clusters
\citep{feretal04}, while the predictions based on the Padova tracks
display a severe discrepancy with the data.  \\
{\bf 4950, z=1.55}. The \mguv\ line was included in the \chiquad\. The
best-fit solution is 3.5 Gyr old, is forming stars according to a {\it
tau} model with e-folding time of 1 Gyr, the metallicity is half-solar
and $E(B-V)=0.15$ (SMC law). The extended star formation history is
consistent with the irregularities and blobs in the $B$-band image and
with its spiral-like morphology (D05).  This galaxy has a huge De
Vauculeurs bulge and turned out to be the most massive object of the
sample. The observed \mguv\ is remarkably well reproduced by this
model. $M_{*}$ is large and very well constrained. It is consistent
with both the value of D05 and that derived here with the BC03
models. The best solution obtained with the BC03 models has similar
overall parameters, but the reddening is larger and the \chiquad\ is
worse.  \\
{\bf 1025, z=1.73.} The \mguv\ line was included in the
\chiquad\. The best-fit is $\sim$ 1.4 Gyr old, still forming stars at
a very modest rate ($\tau$ model, with $\tau=0.3$ Gyr). The stellar
mass is $0.6~10^{11}\msun$. The solution obtained with the BC03 model
has a lower metallicity and a larger mass.\\
{\bf 3523, z=1.76}. The \mguv\ line was not included in the
\chiquad\. The \mguv\ for this galaxy is very small ($0.094 \pm 0.09$
\AA), which implies either very young ages or very low
metallicities. The solution allowing for the best comparison to the
observed line turned out to be young and rather metal-poor. Its \mguv\
is 1.07. The galaxy is passively evolving since 0.2 Gyr. The solution
obtained with the BC03 models allows a worse match to the \mguv\ line
(1.14 \AA) because of the higher metallicity. A higher metallicity is
found because it provides more flux to the near-IR.\\
{\bf 3650, z=1.91}. The \mguv\ line was included in the \chiquad\. The
galaxy is young and metal-rich, and in passive evolution for the M05
models. It has formed stars at a remarkable rate as $\mstar$ is
$0.83\cdot 10^{11}~\msun$. The spectrum displays a nice near-IR excess
that is very well fitted. The match to \mguv\ is remarkable good (1.25
\AA\ versus the observed 1.25 \AA). The solution obtained with the
BC03 models is older and significantly more metal-poor, though the
stellar mass of the galaxy if higher ($2\cdot 10^{11}~\msun$). This
mass is too large for a metallicity $\zsun/5$ in comparison with what
found in the local universe. This solution matches worse to the
observed SED. \\
{\bf 3574, z=1.98}. The \mguv\ line was included in the
\chiquad\. The best-fit is young, metal-rich, and rather highly
reddened. The model is passively evolving. The best-solution obtained
with the BC03 models is older and metal-poor. \\
{\bf 1446, z=2.47}. The \mguv\ line was not included in the
\chiquad\. This is perhaps the strongest case of AGB-dominated
object. The fit requires a super-solar metallicity. However, the
amount of TP-AGB in the M05 models at this metallicity was even
slightly too low around the J band with respect to the
observations. Since the TP-AGB fuel and spectral-types at super-solar
metallicities are not well constrained given the absence of local GCs
as calibrators (see discussion in M05), we could accept to modify the
model. An excellent fit is obtained when the fuel is augmented by 30\%
and the O-rich stars have a slightly warmer temperature. Both changes
are certainly inside the uncertainties of the TP-AGB phase at
high-metallicity. Noteworthy, this fit does not require any
reddening. The stellar mass is less than 1/3 the lower bound of the
D05 solution. The BC03-based solution is older and highly reddened,
yet the match to the observed SED is not satisfactory, in the near-IR
as well as in the optical. We have verified that older ages (2.6 Gyr)
would provide a better fit, but these ages are above the limit set by
the age of the Universe.

Note that the visual quality of the fit did not correspond to
the \chiquad\ values when the F110 band was included in the
\chiquad\. This is due to this photometric data having a very small
error, which drives the \chiquad\. While the best-fit model was found
using all bands for consistency with the other galaxies, the \chiquad\
values given in Table~2 were computed by excluding the F110 band.

The data in F110 and J bands are off the models and are responsible
for the significantly higher values of \chiquad that are obtained in
case of this galaxy with respect to the other objects. We have noted
that significantly better values of \chiquad ($\sim 1.5$) are obtained
when the redshift is higher than the spectroscopic value and close to
the redshift estimated photometrically with the M05 models (2.74,
cf. Figure~\ref{photoz}). In this case, the best-fit model, that has
similar overall parameters as that presented in Figure~\ref{fitred},
did not require any modification of the fuel consumption in the TP-AGB
phase.
\begin{deluxetable}{rccccccc}
\tablewidth{0pt}
\tabletypesize{\scriptsize}
\tablecaption{Photometric data of HUDF galaxies}
\tablehead{
\colhead{ID} &
\colhead{$F435$} &
\colhead{$V$} &
\colhead{$F606$} &
\colhead{$R$} &
\colhead{$F775$} &
\colhead{$F895$} &
\colhead{$F110$} 
}
\tabletypesize{\scriptsize}
\startdata
 8238 & $29.15\pm0.62$ & $28.59\pm1.10$ & $27.42\pm0.09$ & $26.86\pm0.2$ & $25.60\pm0.02$ & $24.66\pm 0.01$ & $23.67\pm0.01$ \\
 4650 & $25.30\pm0.04$ & $24.83\pm0.04$ & $24.78\pm0.02$ & $24.47\pm0.03$ & $23.89\pm0.01$ & $23.18\pm0.01$ & $99$ \\
 1025 & $26.35\pm0.05$ & $25.66\pm0.20$ & $25.84\pm0.02$ & $25.56\pm0.09$ & $25.17\pm0.01$ & $24.52\pm0.01$ & $99$ \\
 3523 & $28.40\pm0.39$ & $28.33\pm0.88$ & $26.77\pm0.06$ & $26.52\pm0.14$ & $25.73\pm0.03$ & $24.99\pm0.02$ & $23.60\pm0.01$ \\
 3650 & $28.51\pm0.31$ & $27.00\pm0.27$ & $26.42\pm0.03$ & $26.10\pm0.10$ & $25.48\pm0.01$ & $24.57\pm0.01$ & $23.36\pm0.01$ \\
 3574 & $28.81\pm0.34$ & $28.49\pm1.04$ & $27.10\pm0.05$ & $27.31\pm0.29$ & $26.38\pm0.03$ & $25.57\pm0.02$ & $24.41\pm0.01$ \\
 1446 & $29.92\pm1.10$  & $28.20\pm0.78$ & $27.52\pm0.08$ & $26.63\pm0.12$ & $26.44\pm0.03$ & $26.12\pm0.04$ & $25.43\pm0.03$ \\
\enddata
\end{deluxetable}

\setcounter{table}{2}
\begin{deluxetable}{cccccccc}
\tablewidth{0pt}
\tabletypesize{\scriptsize}
\tablecaption{{\it Continued}}
\tablehead{
\colhead{$J$} &
\colhead{$F160W$} &
\colhead{$K$} &
\colhead{m($3.6\mu m$)} &
\colhead{m($4.5\mu m$)} &
\colhead{m($5.8\mu m$)} &
\colhead{m($8.0\mu m$)} &
\colhead{$\rm MIPS(24\mu m)$}}
\tabletypesize{\scriptsize}
\startdata
 $23.16\pm0.04$ & $22.76\pm 0.0$ & $22.08\pm0.02$ & $21.29\pm0.13$ & $21.38\pm0.13$ & $21.71\pm0.13$ & $22.06\pm0.15$ & $4.9\pm3$ \\
 $21.57\pm0.02$ & $99$ & $20.54\pm0.11$ & $19.89\pm0.11$ & $19.84\pm0.11$ & $20.11\pm0.11$ & $20.58\pm0.11$ & $26.7\pm4.8$ \\
 $22.74\pm0.07$ & $99$ & $21.80\pm0.05$ & $21.28\pm0.12$ & $21.20\pm0.12$ & $21.27\pm0.12$ & $21.62\pm0.13$ & $31.4\pm3.5$ \\
 $23.19\pm0.04$ & $22.68\pm0.01$ & $22.33\pm0.03$ & $21.60\pm0.14$ & $21.56\pm0.14$ & $21.69\pm0.15$ & $21.88\pm0.15$ & $6.5\pm5.8$ \\
 $22.66\pm0.02$ & $22.28\pm0.00$ & $21.75\pm0.02$ & $21.05\pm0.12$ & $21.01\pm0.12$ & $21.10\pm0.12$ & $21.65\pm0.12$ & $5.8\pm4.3$ \\
 $23.63\pm0.05$ & $23.22\pm0.01$ & $22.67\pm0.03$ & $22.24\pm0.15$ & $22.19\pm0.15$ & $22.12\pm0.15$ & $22.73\pm0.18$ & $15\pm6$ \\
 $25.25\pm0.22$ & $23.38\pm0.01$ & $22.86\pm0.04$ & $22.18\pm0.19$ & $22.08\pm0.19$ & $22.01\pm0.19$ & $22.03\pm0.19$ & $3\pm4.6$ \\
\enddata
\tablenotetext{1.} {The magnitudes are in AB system, which are related to flux
density $f_\nu$ (in erg\ s$^{-1}$\ cm$^{-2}$\ Hz$^{-1}$) by
$m=-2.5\times lg(f_\nu) - 48.60$. Fluxes at 24 $\mu$ are in \microjan}
\tablenotetext{2.} {The reported photometric errors of the IRAC bands reflect the random errors only. Typical systematic errors in these bands are at $\lapprox
0.1$ mag level.}
\end{deluxetable}
\end{document}